\documentclass[aps,pre,twocolumn,groupedaddress,showpacs]{revtex4-1}

\usepackage{amsmath,amssymb,bm}

\begin{document}

\title{Sum rule for response function in nonequilibrium Langevin systems}

\author{Tatsuro Yuge}
\email[]{yuge@m.tains.tohoku.ac.jp}
\affiliation{IIAIR, Tohoku University, Aoba-ku, Sendai 980-8578, Japan}

\date{\today}

\begin{abstract}
We derive general properties of the linear response functions of 
nonequilibrium steady states in Langevin systems.
These correspond to extension of the results
which were recently found in Hamiltonian systems 
[A. Shimizu and T. Yuge, J. Phys. Soc. Jpn. {\bf 79}, 013002 (2010)].
We discuss one of the properties, the sum rule for the response function, 
in particular detail.
We show that the sum rule for the response function of the velocity 
holds in the underdamped case, whereas it is violated in the overdamped case.
This implies that the overdamped Langevin models should be used with great care.
We also investigate the relation of the sum rule to an equality 
on the energy dissipation in nonequilibrium Langevin systems, 
which was derived by Harada and Sasa.
\end{abstract}

\pacs{05.70.Ln, 05.40.-a}

\maketitle

\section{Introduction}

Responses to weak perturbations contain important information 
on the physical properties of systems, 
and are often used to characterize their states.
General properties of the response function are useful 
in investigating the consistency of obtained results both in theories and in experiments, 
and in making prediction from them.
For equilibrium states many general properties are well-known, 
which are based on the linear response theory \cite{KuboTodaHashitsume}.
Much interest has been devoted recently to extending the properties to 
nonequilibrium steady states (NESSs) 
\cite{PhysRevLett.95.130602, PhysRevE.73.026131, EurophysLett.74.391,  JStatPhys.131.543,PhysRevLett.103.090601, PhysRevLett.103.010602, JStatPhys.137.1094, JStatPhys.139.492, JStatMechP07024, PhysRevLett.102.135701, JPSJ.79.013002}.
Several properties of the linear response function of NESSs  
were found by Shimizu and the author 
in general Hamiltonian systems \cite{JPSJ.79.013002}.
These properties consist of experimentally measurable quantities only, 
which is a distinguished feature from the formal results in many other works.
One of the general properties is the sum rule for the response function 
\cite{JPSJ.79.013002}:
\begin{align}
\int_{-\infty}^\infty {\rm Re} \tilde{\Phi}_B^A (\omega ; F) \frac{{\rm d}\omega}{\pi} 
= \bigl\langle \{ B,  A \}_{\rm P} \bigr\rangle_{F,0}, 
\label{introSumRule}
\end{align}
where $A$ is a physical observable, $-B$ is a perturbation potential, and 
$\tilde{\Phi}_B^A$ is the Fourier-Laplace transform of the linear response function of $A$.
$\bigl\langle \cdots \bigr\rangle_{F,0}$ represents the average in a NESS
with a driving force (pump field) $F$ and without the perturbation (probe force), and 
$\{ \bullet,  \bullet \}_{\rm P}$ the Poisson bracket in classical systems 
(or the commutator divided by $i\hbar$ in quantum systems), 
which often reduces to an easily measurable quantity.

In Ref.~\cite{JPSJ.79.013002} they derived these properties 
assuming that a large system (which is composed of the system of interest, 
its environments and a driving source) is a (deterministic) Hamiltonian system.
The applicable range of the properties 
is quite large since almost no assumptions were imposed.
It is not trivial, however, that the properties hold in stochastic systems.
There are two points to examine the validation in such systems.
One is that explicit derivation of the properties in stochastic systems 
extends the universality of the properties.
The other is as follows. 
Some classes of stochastic systems may be regarded 
as approximate models of certain large Hamiltonian systems, 
where the degrees of freedom in the environments are eliminated
(although in some cases 
the connections between these models are not sufficiently clear).
Therefore the validation is utilized as 
a criterion for effectiveness of the approximate models 
because the properties should remain valid 
if the approximations in reducing the original Hamiltonian models to 
the stochastic models are good.

In this paper we show that the sum rule 
holds in the stochastic systems described by the Langevin equations.
It is valid in highly nonequilibrium steady states as well as in equilibrium states.
Here, `highly nonequilibrium' is used in the following two senses: 
(1) the driving force may be arbitrarily large, and 
(2) the intensity of the Langevin noise 
is allowed not to satisfy the second fluctuation-dissipation relation 
(as in the case of nonequilibrium process in laser physics \cite{PhysRevA.57.3074}).
We also derive the asymptotic behavior, 
another general property shown in Hamiltonian systems \cite{JPSJ.79.013002}, 
of response function in the Langevin systems.
Furthermore we show that 
the sum rule for the velocity response function is violated in the overdamped cases.
This implies that the overdamped model sometimes gives results 
inconsistent with those in larger Hamiltonian systems especially when 
they are related to small-time-scale phenomena 
(which are correctly described by the Hamiltonian systems).
We also discuss a relation between the sum rule and the Harada-Sasa equality 
\cite{PhysRevLett.95.130602, PhysRevE.73.026131}, 
which is an expression of the energy dissipation in the Langevin models.

\section{Model\label{sec:model}}

We define a general form of the Langevin model.
The examples are shown in Sec.~\ref{sec:examples}.

We consider a system the state of which is specified by a set of 
$n$ stochastic variables denoted by $\bm{\zeta} = (\zeta_1, \zeta_2, \cdots , \zeta_n)$ 
(called system variables hereafter). 
The dynamics of the system is assumed to be characterized 
by the following stochastic differential equation 
\cite{Gardiner} of a general form:
\begin{align}
{\rm d}\zeta_i (t) &= M_i \bigl(\bm{\zeta}(t) ; F \bigr) {\rm d}t 
+ \sum_{j=1}^n \mathcal{N}_{ij}\bigl(\bm{\zeta}(t) ; F \bigr) \cdot {\rm d}W_j (t) 
\notag\\
&+ \varepsilon f_i^{\rm p}(t) K_i \bigl(\bm{\zeta}(t) \bigr) {\rm d}t.
\label{generalEquation}
\end{align}
The first two terms in the right-hand side describe 
the unperturbed dynamics of the system.
$F$ represents a pump field, 
which determines the degree of nonequilibrium of the system
(e.g., an external driving force or a temperature difference).
The first term is a deterministic part 
($M_i$ is a certain known function of $\bm{\zeta}$ and $F$).
The second one represents the noise term, 
where $W_i(t)$ is a Wiener process.
We assume that the mean of $W_i(t)$ is zero for any $i$ and that 
$W_i(t)$ and $W_j(t)$ are uncorrelated if $i \neq j$.
The symbol `` $\cdot$ '' implies the multiplication in the sense of It\^o.
The noise intensity is determined by $\mathcal{N}_{ij}$ 
(which is a known function of $\bm{\zeta}$ and $F$). 
The third term represents the probe force (perturbation) 
the response to which is of our interest 
($f_i^{\rm p}$ and $K_i$ are known functions of $t$ and of $\bm{\zeta}$, respectively).
We assume that in the unperturbed system (i.e., in the case of $\varepsilon = 0$) 
is realized a certain steady state 
which is stable against perturbations.

A situation to be supposed is as follows.
The system of our interest is driven to a NESS by the pump field $F$.
The steady state may be in a nonlinear response regime, in a linear response regime, 
or an equilibrium state, because $F$ is allowed to be arbitrary (may be very large) 
and the noise intensity is arbitrary. 
To measure the response of the NESS, 
the probe force $\varepsilon f_j^{\rm p} K_j$ is applied to the system in addition to $F$.
(This is sometimes called pump-probe experiment.)
The linear response function to the perturbation $\varepsilon f_j^{\rm p} K_j$ is defined by 
\begin{align}
\Phi_{K_j}^A (t-s; F) 
&= \frac{\delta \left\langle A\bigl(\bm{\zeta}(t) \bigr) \right\rangle_{F, \varepsilon}}
{\varepsilon \delta f_j^{\rm p}(s)} \Bigg|_{\varepsilon=0}. 
\label{definitionResponse}
\end{align}
Here $A$ is a physical observable 
which is a function of the system variables $\bm{\zeta}$, and
$\bigl\langle \cdots \bigr\rangle_{F, \varepsilon}$ represents the average in a state 
with the parameter $F$ and the perturbation.
It should be noted that the response function thus defined is that of a NESS 
which may be in a nonlinear response regime 
because $F$ is allowed to be large (irrespective of $\varepsilon$).
[The meaning of ``linear'' in the linear response function of the NESS 
is the linear order in $\varepsilon f_j^{\rm p} K_j$ (not $F$).]

\section{General Result\label{sec:result}}

\subsection{Sum rule for response function\label{sec:sumRule}}

We now show the sum rule for the response function in the Langevin system.
For this purpose we consider the stochastic differential equation for the observable $A$.
This is given by the It\^o formula \cite{Gardiner}: 
\begin{align}
{\rm d}A\bigl(\bm{\zeta}(t) \bigr) 
&= \Hat{\varLambda} A\bigl(\bm{\zeta}(t) \bigr) {\rm d}t
\notag\\
&+ \sum_{i=1}^n \sum_{j=1}^n \mathcal{N}_{ij}\bigl(\bm{\zeta}(t) ; F \bigr) 
\frac{\partial A}{\partial \zeta_i} \bigl(\bm{\zeta}(t) \bigr) \cdot {\rm d}W_j (t) 
\notag\\
&+ \varepsilon \sum_{i=1}^n f_i^{\rm p}(t) K_i \bigl(\bm{\zeta}(t) \bigr) 
\frac{\partial A}{\partial \zeta_i} \bigl(\bm{\zeta}(t) \bigr) {\rm d}t,
\label{itoFormula}
\end{align}
where the backward operator $\Hat{\varLambda}$ is defined by 
\begin{align}
\Hat{\varLambda} &= 
\sum_{i=1}^n M_i \bigl(\bm{\zeta} ; F \bigr) \frac{\partial}{\partial \zeta_i}
\notag\\
&+ \frac{1}{2} \sum_{i=1}^n \sum_{j=1}^n \sum_{k=1}^n 
\mathcal{N}_{ik}\bigl(\bm{\zeta} ; F \bigr) \mathcal{N}_{jk}\bigl(\bm{\zeta} ; F \bigr) 
\frac{\partial^2}{\partial \zeta_i \partial \zeta_j}.
\label{backwardKramers}
\end{align}

Employing a method similar to that in Ref.~\cite{PhysRevE.73.026131}, 
we transform this differential equation into an integral form:
\begin{align}
&A\bigl(\bm{\zeta}(t) \bigr) = e^{(t-t_0)\Hat{\varLambda}} A\bigl(\bm{\zeta}(t_0) \bigr)
\notag\\
&+ \sum_{i=1}^n \sum_{j=1}^n \int_{t_0}^t \mathcal{N}_{ij}\bigl(\bm{\zeta}(t') ; F \bigr) 
\frac{\partial}{\partial \zeta_i} e^{(t-t')\Hat{\varLambda}} A \bigl(\bm{\zeta}(t') \bigr) 
\cdot {\rm d}W_j (t') 
\notag\\
&+ \varepsilon \sum_{i=1}^n \int_{t_0}^t f_i^{\rm p}(t') K_i \bigl(\bm{\zeta}(t') \bigr) 
\frac{\partial}{\partial \zeta_i} e^{(t-t')\Hat{\varLambda}} 
A \bigl(\bm{\zeta}(t') \bigr) {\rm d}t', 
\label{intEq}
\end{align}
where $t_0$ is an initial time. 
For completeness we provide the detail of the transformation 
in the Appendix.

The average of the second term in the right-hand side of Eq.~(\ref{intEq}) vanishes 
because the integrand is a non-anticipating function 
(the mean value formula of the It\^o stochastic integral) \cite{Gardiner}.
Taking the average of Eq.~(\ref{intEq}), we thus obtain 
\begin{align}
&\left\langle A\bigl(\bm{\zeta}(t) \bigr) \right\rangle_{F, \varepsilon}
= e^{(t-t_0)\Hat{\varLambda}} A\bigl(\bm{\zeta}(t_0) \bigr) 
\notag\\
&+ \varepsilon \sum_{i=1}^n \int_{t_0}^t f_i^{\rm p}(t') 
\left\langle K_i \bigl(\bm{\zeta}(t') \bigr)  
\frac{\partial}{\partial \zeta_i} e^{(t-t')\Hat{\varLambda}} 
A \bigl(\bm{\zeta}(t') \bigr) \right\rangle_{F, \varepsilon} {\rm d}t'.
\label{average}
\end{align}
By a functional differentiation of Eq.~(\ref{average}) with respect to $f_j^{\rm p}(s)$, 
an expression of the response function is derived: 
\begin{align}
\Phi_{K_j}^A (t-s; F) 
= \left\langle K_j \bigl(\bm{\zeta}(s) \bigr) \frac{\partial}{\partial \zeta_j} 
e^{(t-s) \Hat{\varLambda}} A\bigl(\bm{\zeta}(s) \bigr) \right\rangle_{F,0} , 
\label{expression}
\end{align}
for $t>s$ ($>t_0$), 
and $\Phi_{K_j}^A (t-s; F) = 0$ otherwise.
Note that the Langevin system satisfies the causality condition.
If $s\gg t_0$, this expression does not depend on $s$ but on $t-s$ only, 
because a steady state is realized in the unperturbed system.
Thus we finally obtain the sum rule:
\begin{align}
\int_{-\infty}^\infty {\rm Re} \tilde{\Phi}_{K_j}^A (\omega ; F) \frac{{\rm d}\omega}{\pi} 
&= \Phi_{K_j}^A (+0; F) 
\notag\\
&= \left\langle K_j \bigl(\bm{\zeta} \bigr) \frac{\partial A}{\partial \zeta_j} 
\bigl(\bm{\zeta} \bigr) \right\rangle_{F,0}, 
\label{sumRule}
\end{align}
where $\tilde{\Phi}_{K_j}^A (\omega ; F)$ is the Fourier-Laplace transform 
of $\Phi_{K_j}^A (\tau ; F)$ and where we have used 
$\int_{-\infty}^\infty {\rm Im} \tilde{\Phi}_{K_j}^A (\omega ; F) {\rm d}\omega = 0$ 
(this is because $\Phi_{K_j}^A (\tau ; F)$ is a real-valued function).
This is the main result of the present paper.

\subsection{Asymptotic behavior\label{sec:asymptotic}}

In Ref.~\cite{JPSJ.79.013002}, in addition to the sum rule, 
an asymptotic behavior of $\tilde{\Phi}_{K_j}^A (\omega ; F)$ was derived. 
We here show another derivation of it using the sum rule.
We first note that the response function satisfies a dispersion relation
if the system satisfies the causality condition; 
\begin{align}
{\rm Im} \tilde{\Phi}_{K_j}^A (\omega ; F) = 
-\mathcal{P} \int_{-\infty}^\infty \frac{1}{\omega' - \omega}
{\rm Re} \tilde{\Phi}_{K_j}^A (\omega' ; F) \frac{{\rm d}\omega'}{\pi},
\end{align}
where $\mathcal{P}$ denotes the principal value.
By multiplying $\omega$ to the both sides of this equation 
and by taking the $\omega\to\infty$ limit, 
we obtain the asymptotic behavior:
\begin{align}
\lim_{\omega\to\infty} \omega {\rm Im} \tilde{\Phi}_{K_j}^A (\omega ; F) 
&= \int_{-\infty}^\infty {\rm Re} \tilde{\Phi}_{K_j}^A (\omega' ; F) \frac{{\rm d}\omega'}{\pi}
\notag\\
&= \left\langle K_j \bigl(\bm{\zeta} \bigr) \frac{\partial A}{\partial \zeta_j} 
\bigl(\bm{\zeta} \bigr) \right\rangle_{F,0},
\end{align}
where we have exchanged the integration and the limit procedure in the first equality 
and have used the sum rule in the second equality.
Hence the asymptotic behavior 
is valid in the systems (including the Langevin systems) 
where the sum rule and the dispersion relation hold.
The asymptotic value is the same as the sum value in the sum rule.

\subsection{Remarks}

Here we make remarks on the results.

First, Eq.~(\ref{sumRule}) should be interpreted as a prediction on 
the sum value (integral value) of $\tilde{\Phi}_{K_j}^A (\omega ; F)$ 
at many different values of  $\omega$ 
[therefore Eq.~(\ref{sumRule}) is called sum rule].
$\tilde{\Phi}_{K_j}^A (\omega ; F)$ for each $\omega$ 
is easily measured by experiments or numerical simulations, 
e.g., in the following way: 
apply a sinusoidal probe force with frequency $\omega$ 
[i.e., $f^{\rm p}(t) = \sin\omega t$], 
measure the observable  $A$ for sufficiently long time ($\gg 1/\omega$), 
and calculate the Fourier component $\langle \tilde{A}_\omega \rangle_{F, \varepsilon}$ 
at $\omega$ from the time-series data of $A$ 
($\tilde{\Phi}_{K_j}^A (\omega ; F) = 
\lim_{\varepsilon \to 0} \langle \tilde{A}_\omega \rangle_{F, \varepsilon}/\varepsilon$).
The sum rule [Eq.~(\ref{sumRule})] states that the sum value measured in the system 
with a small perturbation equals the average value of 
$S(\bm{\zeta}) \equiv K_j (\bm{\zeta}) \partial A (\bm{\zeta}) \big/ \partial \zeta_j $ 
measured in the system without the perturbation.
In many cases $S$ is a quantity that is easy to measure.
Therefore it is possible to test the sum rule in experiments and simulations.
(In Ref.~\cite{JPSJ.79.013002} we demonstrated the validity of a sum rule 
by a numerical simulation.)

Equation~(\ref{sumRule}) also seems to be a prediction on the value of 
the response function immediately after the probe force is applied, 
as seen in the middle of Eq.~(\ref{sumRule}).
In contrast to the measurement of $\tilde{\Phi}_{K_j}^A (\omega ; F)$, however,  
it is very difficult to measure $\Phi_{K_j}^A (+0; F)$ 
in experiments and numerical simulations.
Therefore it is more natural to interpret Eq.~(\ref{sumRule}) 
as a sum rule for $\tilde{\Phi}_{K_j}^A (\omega ; F)$.

Second, the sum rule holds for any steady states (if they are stable) 
including the states in a nonlinear response regime (at large $F$) 
as well as an equilibrium state ($F=0$).
Although the {\it form} of the rule is the same for all the states, 
the {\it value} of the sum may vary as $F$ is changed.
This is because the steady state distribution function, 
which appears in averaging $S(\bm{\zeta})$ in Eq.~(\ref{sumRule}), 
is dependent on $F$.

Third, the expression of the response function [Eq.~(\ref{expression})] 
in more concrete examples 
was derived in some literatures \cite{JStatPhys.139.492, PhysRevLett.102.135701}, 
although they did not derive the sum rule from it.
The expression itself is not so useful, 
because it is hard to calculate in analytic way 
unless one knows the explicit form of the steady state distribution function, 
and because it is hard to measure in experiments and numerical simulations 
due to the complicated factor $e^{(t-s) \Hat{\varLambda}}$.
In contrast, the rightmost side of Eq.~(\ref{sumRule}) is much easier to measure 
since it does not contain such factors, 
and tests of the validity of Eq.~(\ref{sumRule}) are possible 
as mentioned in the first remark.

Fourth, the sum rule is different from the moment sum rule 
(e.g., Ref.~\cite{JStatPhys.131.543}).
The statement of the moment sum rule is 
that the sum value of $\tilde{\Phi}_{K_j}^A (\omega ; F)$ 
[and $\omega^\lambda \tilde{\Phi}_{K_j}^A (\omega ; F)$]
converges to a certain value. 
However, the dependence of this value on $F$ and the other parameters 
is not given by it, 
although the dependence on $F$ is most interesting point 
in nonequilibrium statistical mechanics. 
In contrast, the statement of the sum rule is not only the convergence of the sum value 
but also its equivalence to the average value of $S(\bm{\zeta})$, 
which gives the dependence of the sum on several parameters 
(including $F$).

$S$ is independent of the state and the system (without probe).
(It depends on the probe force and the observable of our interest, 
both of which may be chosen irrespective of the state and the system.)
Therefore there are no differences in $S$ between equilibrium states and NESSs 
and between non-interacting systems and interacting systems.
The dependence on the state and the system appears only in averaging it, 
which leads to the dependence of the sum on $F$ in general.
When $S$ is independent of $\bm{\zeta}$, in particular, 
this dependence disappears and 
the sums are the same for any steady states and for any systems.
This fact cannot be predicted by the moment sum rule.

\section{Examples\label{sec:examples}}

In this section we describe some examples of the Langevin model defined 
in the general form [Eq.~(\ref{generalEquation})] in Sec.~\ref{sec:model}, 
and see the concrete forms of the sum rule in the examples.
It should be noted that in Eq.~(\ref{generalEquation}) 
the relevant quantities to the sum rule are $K_i$'s.

Furthermore we show that the sum rule for the velocity response function 
is violated in the overdamped case.

\subsection{Underdamped case\label{sec:underdamped}}

One of the simplest examples is the single-particle underdamped Langevin model 
in one dimension which is described by the following equations: 
\begin{align}
\frac{{\rm d}p(t)}{{\rm d}t} = &-\frac{\gamma}{m}p(t) 
- \frac{\partial U}{\partial x}\bigl(x(t)\bigr) + F +\xi(t) 
\notag\\
&+ \varepsilon f^{\rm p}(t) \frac{\partial B}{\partial x}\bigl(x(t)\bigr),
\label{underdamped1}
\\
\frac{{\rm d}x(t)}{{\rm d}t} = &~~ \frac{p(t)}{m}.
\label{underdamped2}
\end{align}
Here, $p$, $x$, $m$, and $\gamma$ are 
the momentum, position, mass and friction coefficient of the particle, respectively.
$U(x)$ is a potential, $F$ is an external driving force, 
and $-B(x)$ is a perturbation potential of the probe.
$\xi(t)$ is a white Gaussian noise with zero mean and satisfies 
$\langle \xi(t) \xi(t')\rangle = 2D\delta(t-t')$.
$F$ is allowed to be arbitrarily strong.
A nonequilibrium steady state in the nonlinear response regime is realized for large $F$, 
while an equilibrium state is realized for $F=0$.

More precise forms of  Eqs.~(\ref{underdamped1}) 
and (\ref{underdamped2}) are given by Eq.~(\ref{generalEquation})
where $n=2$, $\zeta_1=p$, and $\zeta_2=x$,  and where 
\begin{align}
M_1 (p,x; F) &= -\frac{\gamma}{m}p - \frac{\partial U}{\partial x}(x) + F, 
~~
M_2 (p,x; F) = \frac{p}{m},
\notag\\
\mathcal{N}_{11} (p,x; F) &= \sqrt{2D}, 
~~
\mathcal{N}_{ij \neq 11} (p,x; F) = 0, 
\notag\\
K_1 (p,x) &= \frac{\partial B}{\partial x}(x), 
~~
K_2 (p,x) = 0.
\notag
\end{align}
In this case the sum rule reads
\begin{align}
\int_{-\infty}^\infty {\rm Re} \tilde{\Phi}_B^A (\omega ; F) \frac{{\rm d}\omega}{\pi} 
= \left\langle \frac{\partial B}{\partial x}(x) 
\frac{\partial A}{\partial p}(p,x) \right\rangle_{F,0}.
\label{sumRuleInSingleParticle0}
\end{align}
This form is the same as that in the classical Hamiltonian models 
where the systems of interest are single-particle systems.
In particular, when considering the momentum response 
to the spatially homogeneous probe force (i.e., $A=p$ and $B=x$), 
we have the $F$-independent sum value:
\begin{align}
\int_{-\infty}^\infty {\rm Re} \tilde{\Phi}_x^p (\omega ; F) \frac{{\rm d}\omega}{\pi} = 1.
\label{sumRuleInSingleParticle}
\end{align}

The second example is a three-dimensional many-particle 
underdamped Langevin model described by
\begin{align}
\frac{{\rm d}\vec{p}_\mu(t)}{{\rm d}t} = &-\frac{\gamma_\mu}{m_\mu}\vec{p}_\mu(t) 
- \frac{\partial U}{\partial \vec{r}}\bigl(\vec{r}_\mu(t)\bigr) 
- \frac{\partial V}{\partial \vec{r}_\mu}\bigl(\{\vec{r}_\nu(t)\}\bigr) 
\notag\\
&+ \vec{F} + \vec{\xi}_\mu(t) 
+ \varepsilon f^{\rm p}(t) \frac{\partial B}{\partial \vec{r}}
\bigl( \vec{p}_\mu(t), \vec{r}_\mu(t) \bigr),
\label{manyunderdamped1}
\\
\frac{{\rm d}\vec{r}_\mu(t)}{{\rm d}t} = &~\frac{\vec{p}_\mu(t)}{m_\mu} 
- \varepsilon f^{\rm p}(t) \frac{\partial B}{\partial \vec{p}}
\bigl( \vec{p}_\mu(t), \vec{r}_\mu(t) \bigr),
\label{manyunderdamped2}
\end{align}
for $\mu=1, 2, \cdots , N$.
Here, $\vec{a} = (a^1, a^2, a^3)$ represents a three-dimensional vector.
$\vec{p}_\mu$, $\vec{r}_\mu$, $m_\mu$, and $\gamma_\mu$ are the momentum, 
position, mass and friction coefficient of the $\mu$th particle, respectively.
$U(\vec{r})$ is a single-particle potential 
and $V(\{\vec{r}_\nu\})$ is an inter-particle potential.
$\vec{F}$ is an external driving force, 
the strength of which is arbitrary.
$-B(\vec{p}, \vec{r})$ is a probe potential (perturbation) 
which is assumed to depend on a momentum as well as on a position.
(An example is an interaction of an electron with an electromagnetic field.)
$\vec{\xi}_\mu(t)$ is a white Gaussian noise with zero mean and satisfies 
$\langle \xi_\mu^\alpha(t) \xi_{\mu'}^{\alpha'}(t')\rangle 
= 2D \delta_{\alpha\alpha'} \delta_{\mu\mu'} \delta(t-t')$ $(\alpha ,\alpha' = 1, 2, 3)$.
More precise forms of these equations 
are given by Eq.~(\ref{generalEquation}) with $n=6N$, 
$\zeta_{6\mu-6+\alpha} = p_\mu^\alpha$, and 
$\zeta_{6\mu-3+\alpha} = r_\mu^\alpha$ $(\alpha = 1, 2, 3)$.
The relevant quantities $K_i$'s to the sum rule are written as 
\begin{align}
K_{6\mu-6+\alpha}\bigl( \{\vec{p}_\nu, \vec{r}_\nu\} \bigr) 
&= \frac{\partial B}{\partial r^\alpha}(\vec{p}_\mu, \vec{r}_\mu),~
\notag\\
K_{6\mu-3+\alpha}\bigl( \{\vec{p}_\nu, \vec{r}_\nu\} \bigr) 
&= \frac{\partial B}{\partial p^\alpha}(\vec{p}_\mu, \vec{r}_\mu).
\notag
\end{align}
We thus obtain the sum rule for the response function 
$\Phi_B^A (t-s; F) = 
\delta \bigl\langle A\bigl( \{\vec{p}_\mu(t), \vec{r}_\mu(t)\} \bigr)\bigr\rangle_{F,0} 
\big/ \varepsilon \delta f^{\rm p}(s) \big|_{\varepsilon =0}$
\begin{align}
\int_{-\infty}^\infty & {\rm Re} \tilde{\Phi}_B^A (\omega ; F) \frac{{\rm d}\omega}{\pi} 
\notag\\
= & \left\langle \sum_{\mu,\alpha} \left[
\frac{\partial B}{\partial r_\mu^\alpha}
\bigl( \{\vec{p}_\nu, \vec{r}_\nu\} \bigr) 
\frac{\partial A}{\partial p_\mu^\alpha}
\bigl( \{\vec{p}_\nu, \vec{r}_\nu \} \bigr) \right. \right.
\notag\\
& - \left. \left. \frac{\partial B}{\partial p_\mu^\alpha}
\bigl( \{\vec{p}_\nu, \vec{r}_\nu\} \bigr) 
\frac{\partial A}{\partial r_\mu^\alpha}
\bigl( \{\vec{p}_\nu, \vec{r}_\nu \} \bigr) \right] \right\rangle_{F,0}.
\end{align}
This form is the same as that in the classical Hamiltonian models 
[Eq.~(\ref{introSumRule})].
This reduces to Eq.~(\ref{sumRuleInSingleParticle0}) 
if $N=1$, $\partial B / \partial p = 0$ and the motion is restricted to one dimension.

In Langevin models the noise intensity $D$ is usually assumed 
to be equal to $\gamma k_{\rm B}T$ 
[which is called the second fluctuation-dissipation relation (FDR) 
\cite{KuboTodaHashitsume}], where $T$ is the temperature of the environment 
and $k_{\rm B}$ is the Boltzmann constant.
However, as seen in the general result and in the above two examples, 
this assumption is not necessary for the validity of the sum rule 
because the noise intensity does not explicitly contribute to the rule.
This indicates that the sum rule holds in a very wide range of systems 
from highly nonequilibrium systems (in the sense that the second FDR is violated) 
to purely deterministic systems ($D=0$).
The violation of the second FDR is often seen 
in the treatment of nonequilibrium process (far from equilibrium) 
in light-emitting devices by quantum Langevin equation \cite{PhysRevA.57.3074}.

For the same reason the noise intensity $D$ may depend on $F$.
Therefore the treatment in the present paper is applicable also to 
the systems in which the second FDR is gradually violated as $F$ increases.
Also $D$ may depend on the position of the particles 
in the two examples and on $\mu$ and $\alpha$ in the second example.
One of the consequences from this fact is that 
the sum rule is valid also in heat conducting nonequilibrium systems
driven by temperature difference.
It should be noted that the {\it validity} of the sum rule is independent 
of the noise intensity 
whereas the {\it value} of the sum depends on it 
because the steady states depend on it in general.

Finally we make a brief comment.
In some Langevin systems the natural interpretation of the Langevin equations 
used in physics is the Stratonovich-type stochastic differential equations 
\cite{bookSekimoto}.
In such cases we should first interpret the given Langevin equations to 
the corresponding Stratonovich-type equations, then transform them 
into the It\^o-type equations [Eq.~(\ref{generalEquation})], 
and finally apply the general form of the sum rule.
Note that the {\it form} of the sum rule is identical 
irrespective of the interpretations, 
although the {\it value} of the sum depends on them 
because the steady states may be different in different interpretations.
In the above examples (with position-independent $D$), 
the values as well as the forms of the sum rules are independent of the interpretations 
because the forms of the Stratonovich-type and It\^o-type equations 
are the same except for the senses of the multiplications in these cases.

\subsection{Overdamped case\label{sec:overdamped}}

In this subsection we discuss the sum rule 
in the overdamped case.
For simplicity we consider a single-particle model 
(the extension to many-particle models is straightforward).
We consider the single-particle overdamped Langevin model in one dimension 
which is described by 
\begin{align}
\gamma \frac{{\rm d}x(t)}{{\rm d}t} = 
- \frac{\partial U}{\partial x}\bigl(x(t)\bigr) + F +\xi(t) 
+ \varepsilon f^{\rm p}(t) \frac{\partial B}{\partial x}\bigl(x(t)\bigr).
\label{overdamped}
\end{align}
Here the notations are the same as those 
in Eqs.~(\ref{underdamped1}) and (\ref{underdamped2}).
$\xi(t)$ is a white Gaussian noise with zero mean and satisfies 
$\langle \xi(t) \xi(t')\rangle = 2D\delta(t-t')$.
A more precise form of  Eq.~(\ref{overdamped}) 
is given by Eq.~(\ref{generalEquation})
where $n=1$ and $\zeta_1=x$,  and where 
\begin{align}
M_1 (x; F) &=  - \frac{1}{\gamma} \left[ \frac{\partial U}{\partial x}(x) + F \right], 
\notag\\
\mathcal{N}_{11} (x; F) &= \frac{\sqrt{2D}}{\gamma}, 
\notag\\
K_1 (x) &= \frac{1}{\gamma} \frac{\partial B}{\partial x}(x).
\notag
\end{align}
Even in this model the sum rule is valid 
{\it if the observable $A$ is a function of only $x$}; 
\begin{align}
\int_{-\infty}^\infty {\rm Re} \tilde{\Phi}_B^A (\omega ; F) \frac{{\rm d}\omega}{\pi} 
= \frac{1}{\gamma} \left\langle \frac{\partial B}{\partial x}(x) 
\frac{\partial A}{\partial x}(x) \right\rangle_{F,0}.
\end{align}

One sometimes considers as an observable of interest the velocity $v$, 
which is defined by $v(t) {\rm d}t = {\rm d}x(t)$ in the overdamped model.
Because we have not derived the sum rule in the cases that the observable 
is a function of the differentials of the system variables, 
the general result [Eq.~(\ref{sumRule})] does not ensure that the sum rule holds 
for the response function $\Phi_B^v$ of the velocity in this model.
However, from the viewpoint that the overdamped model is regarded as 
a coarse-grained model of the underdamped one, 
$\Phi_B^v$ must satisfy the sum rule 
if the coarse-graining procedure is good.

To investigate this point we again consider the underdamped model  
described by Eqs.~(\ref{underdamped1}) and (\ref{underdamped2}).
Because the momentum of the particle is included in the system variables 
in the underdamped model, we can safely consider the sum rule 
for the response function of the velocity $p/m$, 
which reads 
\begin{align}
\int_{-\infty}^\infty {\rm Re} \tilde{\Phi}_B^{p/m} (\omega ; F) \frac{{\rm d}\omega}{\pi} 
= \frac{\gamma}{m} \frac{1}{\gamma} 
\left\langle \frac{\partial B}{\partial x} \right\rangle_{F,0}.
\label{sumRuleInSingleParticle2}
\end{align}
The right-hand side diverges in the overdamped limit ($m/\gamma \to 0$).
Therefore a necessary condition for the validity of the sum rule for $\Phi_B^v$ 
is that $\Phi_B^v(+0 ; F)$ is divergent in the overdamped model.
We next examine this condition directly in the overdamped model.
$\Phi_B^v$ is calculated by averaging Eq.~(\ref{overdamped}) and 
by functionally differentiating the result with respect to $f(s)$: 
\begin{align}
\Phi_B^v (t-s ; F) &= \frac{\delta}{\varepsilon \delta f(s)} 
\left\langle \frac{{\rm d}x(t)}{{\rm d}t} \right\rangle_{F, \varepsilon} 
\bigg|_{\varepsilon=0}
\notag\\
&= \frac{1}{\gamma^2} \left\langle \frac{\partial B}{\partial x}\bigl(x(s)\bigr) 
\frac{\partial}{\partial x} e^{(t-s) \Hat{\varLambda}} 
\frac{\partial U}{\partial x}\bigl(x(s)\bigr)\right\rangle_{F,0}
\notag\\
&+ \frac{1}{\gamma} \left\langle \frac{\partial B}{\partial x}(x) 
\right\rangle_{F,0} \delta(t-s), 
\label{v_response}
\end{align}
where we have used the expression [Eq.~(\ref{expression})] 
in the first term in the rightmost side.
Owing to the second term, $\Phi_B^v(+0 ; F)$ is divergent \cite{note:overdamped}. 
Thus the sum rule for $\Phi_B^v$ seems to be valid in the overdamped model 
in the sense that the both sides of it are divergent.
More careful analysis, however, reveals that the diverging behaviors are different 
between Eq.~(\ref{v_response}) and 
the overdamped limit of Eq.~(\ref{sumRuleInSingleParticle2}).
The diverging behavior of Eq.~(\ref{v_response})  is dominated by the delta function 
in the second term in the rightmost side.
This comes from the functional derivative $\delta f(t) / \delta f(s)$, 
the order of which is estimated as $O(1/\Delta t)$.
Here $\Delta t$ is the smallest time scale of our observation on the system.
On the other hand, the diverging behavior 
of the overdamped limit of Eq.~(\ref{sumRuleInSingleParticle2}) 
is dominated by the factor $\gamma / m$ in the right-hand side.
Since the overdamped Langevin model should be interpreted 
as an effective description of the underdamped one in the time scale 
$\Delta t \gg m/\gamma$, 
one must first take the $m/\gamma \to 0$ limit and then take the $\Delta t \to 0$ limit 
to have continuous time limit of the overdamped model.
Therefore the diverging behavior is stronger 
in the overdamped limit of Eq.~(\ref{sumRuleInSingleParticle2})  [$O(\gamma / m)$]
than in Eq.~(\ref{v_response}) [$O(1/\Delta t)$].
In this sense the sum rule for $\Phi_B^v$ is violated in the overdamped model.
This is consistent with the fact that the overdamped Langevin model is valid 
(as a coarse-grained model of the underdamped one) 
only in the frequency range of $\omega \ll \gamma / m$ 
(which would result in an incorrect contribution to the sum 
from the higher frequency region).

\section{Relation to the Harada-Sasa equality\label{sec:HaradaSasa}}

In Refs.~\cite{PhysRevLett.95.130602,PhysRevE.73.026131}, 
Harada and Sasa derived an equality (the Harada-Sasa equality) 
on the energy dissipation rate in nonequilibrium Langevin systems.
We here discuss the relation between this equality and the sum rule. 

For simplicity we consider the single-particle underdamped Langevin model 
described by Eqs.~(\ref{underdamped1}) and (\ref{underdamped2}).
In this model the energy dissipation rate $J$ from the system to the environment 
is defined by \cite{JPSJ.66.1234,bookSekimoto}
\begin{align}
J(t) {\rm d}t = \left(\frac{\gamma}{m}p(t) - \xi(t) \right) \circ {\rm d}x(t), 
\label{Sekimoto}
\end{align}
where the symbol `` $\circ$ '' represents the multiplication in the sense of Stratonovich.
The Harada-Sasa equality is an expression of the average of $J$:
\begin{align}
\langle J \rangle_{F,0} &= \frac{\gamma}{m^2} \langle p \rangle_{F,0}{}^2
\notag\\
&+ \frac{\gamma}{m^2} \int_{-\infty}^\infty \left[ \tilde{C}^p (\omega ; F) 
- 2mk_{\rm B} T {\rm Re} \tilde{\Phi}_x^p (\omega ; F) \right] \frac{{\rm d}\omega}{2\pi},
\end{align}
where the usual assumption (the second FDR) $D=\gamma k_{\rm B}T$ 
on the noise intensity is imposed.
$\tilde{C}^p (\omega ; F)$ is the Fourier transform of 
the time-correlation function $C^p(\tau ; F)$ of the momentum; 
$C^p(\tau ; F) = \bigl\langle \bigl( p(\tau) - \langle p \rangle_{F,0} \bigr) 
\bigl( p(0) - \langle p \rangle_{F,0} \bigr) \bigr\rangle_{F,0}$.

We show that this equality is derived with the help of the sum rule.
First we note that Eq.~(\ref{Sekimoto}) is rewritten in the It\^o type:
$J(t) {\rm d}t = \bigl(\gamma p^2(t) / m^2 - D/m \bigr) {\rm d}t
- (\sqrt{2D}/m) p(t) \cdot {\rm d}W(t).$
This is derived by substituting ``$\xi(t) = \sqrt{2D} {\rm d}W(t) / {\rm d}t$'' 
and ${\rm d}x(t)=p(t){\rm d}t/m$ into Eq.~(\ref{Sekimoto}), 
and by using the Stratonovich-It\^o transformation \cite{Gardiner}.
Since the average of the last term vanishes 
due to the mean value formula, 
the average dissipation rate is written in a simple form 
\cite{JStatPhys.21.191,bookSekimoto}:
\begin{align}
\langle J \rangle_{F,0} = \frac{\gamma}{m^2} \langle p^2 \rangle_{F,0} - \frac{D}{m}.
\label{enegyDissipation}
\end{align}
By multiplying 
$\int_{-\infty}^\infty {\rm Re} \tilde{\Phi}_x^p (\omega ; F) {\rm d}\omega /\pi$,
which is equal to 1 owing to the sum rule [see Eq.~(\ref{sumRuleInSingleParticle})], 
to the last term, and by noting $\langle p^2 \rangle_{F,0} = \langle p \rangle_{F,0}{}^2
+ \int_{-\infty}^\infty \tilde{C}^p (\omega ; F) {\rm d}\omega /2\pi$, 
we obtain 
\begin{align}
\langle J \rangle_{F,0} &= \frac{\gamma}{m^2} \langle p \rangle_{F,0}{}^2
\notag\\
&+ \frac{\gamma}{m^2} \int_{-\infty}^\infty \left[ \tilde{C}^p (\omega ; F) 
- \frac{2mD}{\gamma} {\rm Re} \tilde{\Phi}_x^p (\omega ; F) \right] 
\frac{{\rm d}\omega}{2\pi}.
\end{align}
This becomes the Harada-Sasa equality 
if $D=\gamma k_{\rm B}T$ \cite{note:HaradaSasa}.

As seen in the above derivation, the validation of the Harada-Sasa equality 
requires that Eqs.~(\ref{sumRuleInSingleParticle}) and (\ref{enegyDissipation}) hold.
The former holds in a wide range of nonequilibrium systems 
(even in systems other than Langevin systems) 
because it is a specific form of the sum rule.
In contrast, the validity range of the latter is not so large.
The meaning of Eq.~(\ref{enegyDissipation}) with $D=\gamma k_{\rm B}T$ is 
that the ratio of the average dissipation rate and the difference between 
the kinetic energy of the system and the environment temperature,
$\langle J \rangle_{F,0} \big/ \bigl( \langle p^2 \rangle_{F,0} /m - k_{\rm B}T \bigr)$, 
is a constant, $\gamma/m$, which is independent of $F$.
In some systems other than Langevin systems, however, this does not hold, 
especially in states far from equilibrium.
For example, in a numerical simulation of a model of electrical conduction 
\cite{JPSJ.74.1895,PTP.178.64}, it is clearly seen that 
$\langle J \rangle_{F,0} \big/ \bigl( \langle p^2 \rangle_{F,0} /m - k_{\rm B}T \bigr)$ 
does depend on $F$ \cite{note:unpublished}.
In this sense the validity range of the Harada-Sasa equality 
is restricted to the systems in which the ratio is constant.
A sufficient condition for this might be the distinct separation of time scales, 
as they mentioned in Ref.~\cite{PhysRevE.73.026131}.
It should be noted, however, that
even when the ratio is not constant there remains another possibility. 
That is, it might be possible that 
one can define an $F$-dependent $\gamma (F)$ in a certain way 
and that $\gamma (F)/m$ equals the ratio 
$\langle J \rangle_{F,0} \big/ \bigl( \langle p^2 \rangle_{F,0} /m - k_{\rm B}T \bigr)$ 
in the steady state at each $F$.
Whether this is true or not should be tested 
in systems (e.g., in the model in Refs.~\cite{JPSJ.74.1895,PTP.178.64}) 
which cannot be described by Langevin equations.

\section{Summary\label{sec:summary}}

In this paper we extended the validity range of the sum rule 
(and the asymptotic behavior) 
for the linear response function of steady states 
to a class of stochastic models described 
by a general form of the Langevin equations. 
This holds for a wide range of the steady states (if they are stable) 
including highly nonequilibrium states as well as equilibrium states, 
because the driving force is allowed to be large (e.g., to be in a nonlinear response regime)
and the noise intensity may be arbitrary (e.g., not to satisfy 
the second fluctuation-dissipation relation).
The sum rule is a property which normal nonequilibrium models should have 
and therefore is used as a touchstone to examine the correctness of 
results in experiments and theories of NESSs.
In the overdamped Langevin model the sum rule for the velocity 
response function does not hold, which suggests that results in the 
overdamped model should be treated with care if they are concerned 
with small time scales. 

We also showed the relation of the sum rule to the Harada-Sasa equality.
The equality is reduced to a simpler form 
when one uses the sum rule with a specific choice of observable.

Further extension of the validity range of the sum rule remains as theoretical issues.
Extension to time-dependent case where $M_i$ and $\mathcal{N}_{ij}$ 
in Eq.~(\ref{generalEquation}) explicitly depend on $t$ is straightforward.
It would be also interesting to examine the validation of the sum rule 
in non-Markovian models.
It is easily extended to the non-Markovian models which become Markovian 
if appropriate dynamical variables are added to the original system variables.
For other non-Markovian models, the method used 
in Refs.~\cite{PhysRevE.74.026112,JStatMechP10010} 
to generalize the Harada-Sasa equality to non-Markovian cases might give hints.
Finally, it is also important to investigate higher-order responses of NESSs 
in stochastic systems as studied in Hamiltonian systems recently \cite{arXivShimizu}.


\begin{acknowledgments}
The author thanks A. Shimizu for helpful discussions 
and critical reading of the manuscript.
This work was supported by the Grant-in-Aid for the GCOE Program 
``Weaving Science Web beyond Particle-Matter Hierarchy.''
\end{acknowledgments}


\appendix*

\section{Derivation of Eq.~(\ref{intEq})\label{sec:derivation}}

We introduce the time-evolution operator $\Hat{\varOmega}(t)$ 
corresponding to Eq.~(\ref{itoFormula}); 
for an arbitrary function $C$ of $\zeta$, the value at $\zeta(t)$ is given by 
$C\bigl( \zeta(t) \bigr) = \Hat{\varOmega}(t) C\bigl( \zeta(t_0) \bigr)$, 
where $t_0$ is the initial time and $\Hat{\varOmega}(t_0)=1$.
Then Eq.~(\ref{itoFormula}) is rewritten as 
\begin{align}
{\rm d} \Hat{\varOmega}(t) A\bigl(\bm{\zeta}(t_0) \bigr) 
&= \Hat{\varOmega}(t) \left[ \Hat{\varLambda} A \right] 
\bigl(\bm{\zeta}(t_0) \bigr) {\rm d}t
\notag\\
&+ \sum_{i=1}^n \sum_{j=1}^n \Hat{\varOmega}(t) \left[ \mathcal{N}_{ij} 
\frac{\partial A}{\partial \zeta_i} \right] \bigl(\bm{\zeta}(t_0) \bigr) \cdot {\rm d}W_j (t) 
\notag\\
&+ \varepsilon \sum_{i=1}^n f_i^{\rm p}(t) \Hat{\varOmega}(t) \left[ K_i 
\frac{\partial A}{\partial \zeta_i} \right] \bigl(\bm{\zeta}(t_0) \bigr) {\rm d}t.
\label{itoFormula2}
\end{align}
Because $A$ is arbitrary the above equation is regarded as 
a stochastic differential equation for $\Hat{\varOmega}$.
In order to transform this equation into an integral equation, 
we introduce an operator, 
$\Check{\varOmega}(t) = \Hat{\varOmega}(t) e^{-(t-t_0)\Hat{\varLambda}}$.
From Eq.~(\ref{itoFormula2}) the differential equation for $\Check{\varOmega}$ 
is given by 
\begin{align}
{\rm d}\Check{\varOmega}(t) 
&= \sum_{i=1}^n \sum_{j=1}^n\Hat{\varOmega}(t) \left[ \mathcal{N}_{ij} 
\frac{\partial}{\partial \zeta_i} e^{-(t-t_0)\Hat{\varLambda}} \right] \cdot {\rm d}W_j (t) 
\notag\\
&+ \varepsilon \sum_{i=1}^n f_i^{\rm p}(t)\Hat{\varOmega}(t) \left[ K_i 
\frac{\partial}{\partial \zeta_i} e^{-(t-t_0)\Hat{\varLambda}} \right] {\rm d}t.
\label{itoFormula3}
\end{align}
Then formally integrating this equation from $t_0$ to $t$ 
with the initial condition $\Check{\varOmega}(t_0)=1$ 
and multiplying $e^{(t-t_0)\Hat{\varLambda}}$ from the right,
we obtain 
\begin{align}
\Hat{\varOmega}(t) &= e^{(t-t_0)\Hat{\varLambda}}
\notag\\
&+ \sum_{i=1}^n \sum_{j=1}^n \int_{t_0}^t \Hat{\varOmega}(t') \left[ \mathcal{N}_{ij} 
\frac{\partial}{\partial \zeta_i} e^{(t-t')\Hat{\varLambda}} \right] \cdot {\rm d}W_j (t') 
\notag\\
&+ \varepsilon \sum_{i=1}^n \int_{t_0}^t f_i^{\rm p}(t')\Hat{\varOmega}(t') \left[ K_i 
\frac{\partial}{\partial \zeta_i} e^{(t-t')\Hat{\varLambda}} \right] {\rm d}t'.
\end{align}
Acting this equation on $A\bigl( \zeta(t_0) \bigr)$, 
we finally get Eq.~(\ref{intEq}).


\begin{thebibliography}{99}
\bibitem{note:overdamped}
From the viewpoint that the overdamped model is an approximate one of 
a certain Hamiltonian model, the infinite value of the sum 
in the overdamped model implies the violation of the sum rule 
because it must be finite in Hamiltonian systems \cite{JPSJ.79.013002}.
Furthermore, since all the systems in experiments are described 
by Hamiltonian systems 
(if the degrees of freedom in sufficiently large environments are included), 
it should be experimentally verified (e.g., in experiments of the Brownian motion 
of a bead in air \cite{Science.328.1673}) that the overdamped model 
gives an incorrect result about the sum rule.
\bibitem{note:HaradaSasa}
This equivalence was pointed out 
also in their original paper \cite{PhysRevE.73.026131}.
\end{thebibliography}
\end{document}